\documentclass{iaus}
\usepackage{graphicx}

\newcommand\arcdeg{\mbox{$^\circ$}}

\title[JD 11.~~Signpost of Multiple Planets ] 
{Signpost of Multiple Planets in Debris Disks}

\author[Su et al.]   
{Kate Y. L. Su \and G. H. Rieke}

\affiliation{Steward Observatory, University of Arizona, \\
  933 N Cherry Avenune, AZ 85750, USA \\ email: {\tt ksu@as.arizona.edu}, {\tt grieke@as.arizona.edu} }

\pubyear{2013}
\volume{299}  
\pagerange{}
\jname{Exploring the Formation and Evolution of Planetary Systems}
\editors{Brenda Matthews \& James Graham, eds.}
\begin{document}

\maketitle

\begin{abstract}

We review the nearby debris disk structures revealed by
multi-wavelength images from {\it Spitzer} and {\it Herschel}, and
complemented with detailed spectral energy distribution
modeling. Similar to the definition of habitable zones around stars,
debris disk structures should be identified and characterized in terms
of dust temperatures rather than physical distances so that the
heating power of different spectral type of stars is taken into
account and common features in disks can be discussed and compared
directly. Common features, such as warm ($\sim$150 K) dust belts near
the water-ice line and cold ($\sim$50 K) Kuiper-belt analogs, give
rise to our emerging understanding of the levels of order in debris
disk structures and illuminate various processes about the formation
and evolution of exoplanetary systems. In light of the disk structures
in the debris disk twins (Vega and Fomalhaut), and the current limits
on the masses of planetary objects, we suggest that the large gap
between the warm and cold dust belts is the best signpost for multiple
(low-mass) planets beyond the water-ice line.


\end{abstract}

\firstsection 

\vspace{-0.2cm}
\section{Five Zones of Debris Dust}

\begin{figure}[hbt]
\vspace*{-0.5 cm}
\begin{center}
 \includegraphics[width=4.72in]{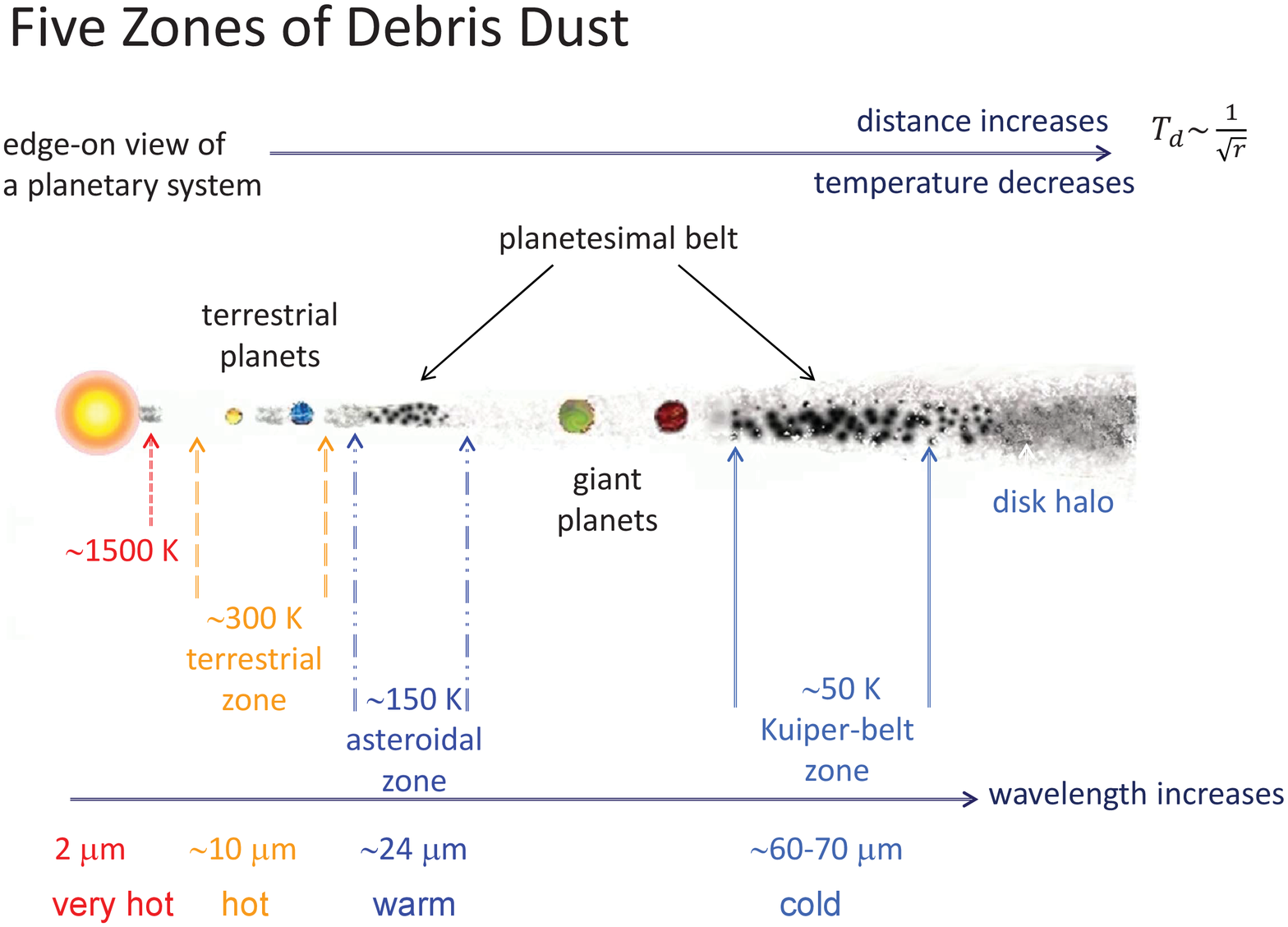} 
 \vspace*{-0.8 cm}
 \caption{Illustration for the five zones of debris dust.}
   \label{fig1}
\end{center}
\end{figure}

Great advances are being made in the observations of debris disks
using all available platforms, revealing enduring patterns in the
behavior of debris disk structure and its evolution that point toward
general aspects of planetary system information and development that
are not accessible by any other means. Although each individual
resolved disk appears to be different from the others, combining all
multi-wavelength observations from many different ground- and
space-based facilities, debris disk structures can be broadly
characterized by five different zones at roughly four different
temperatures with thermal emission peaked at four different
wavelengths (see Fig.\ \ref{fig1}). A majority of disks posses dense,
cold ($\sim$50 K), Kuiper-belt-analog dust that emits prominently in
the far-infrared near 70 $\mu$m. Closer to the star, $\sim$20\% of the
debris systems have a warm ($\sim$150 K), asteroid-like-analog dust
near the water-ice line (Ballering et al.\ 2013). Even closer, dust in
the terrestrial zone has a temperature of $\sim$300 K with emission
peaked around 10 $\mu$m. Some debris systems even have $\sim$1500 K,
very hot excess emission that has been detected through ground-based
near-infrared interferometric techniques. Finally, the fifth zone is a
large halo, an extended structure, surrounding the aforementioned four
zones, that only contains small grains. Not all debris disks have
these five zones. Furthermore, except for the one at the water-ice
line, the temperatures of the zones and hence their spectral energy
distributions (SEDs) depend on the exact locations of the leftover
planetesimal belts and the shepherding planets. Therefore, the
variation in these zones around different stars is probably related to
the various paths by which a system forms and evolves.

Since the majority of debris systems are not spatially resolved, it is
not trivial to convert the observed properties like measured dust
temperatures to physical parameters like the location of planetesimal
belts. The dust temperatures can be well constrained from well-sampled
(both Wien- and Rayleigh-Jeans sides of the emission) 
disk SEDs from mid- to far-infrared. With an additional assumption
that grains in a debris disk are generated in collisional cascades
that generally result in a steep size distribution (Gaspar et al.\
2012), the observed (optically thin) disk emission is sensitive to the
opacity of the disk, which is dominated by small grains. Ideally we
can infer the location of the dust from the observed dust temperature
and the average grain size in a disk and its optical properties. Using
{\it Spitzer} IRS spectra and broad-band 70/100 $\mu$m photometry from
{\it Spitzer} and {\it Herschel}, roughly a quarter of debris systems
require two (warm and cold) temperatures to fit the disk
emission. Although not all these two-temperature debris systems have
two distinct dust belts, the ones that have resolved disk images at
multiple wavelengths strongly suggest two separate planetesimal belts
as we discuss below.


\vspace{-0.3cm}
\section{Nearby Resolved Debris Disk Structures}

\begin{figure}[thb]
\begin{center}
 \includegraphics[width=\linewidth]{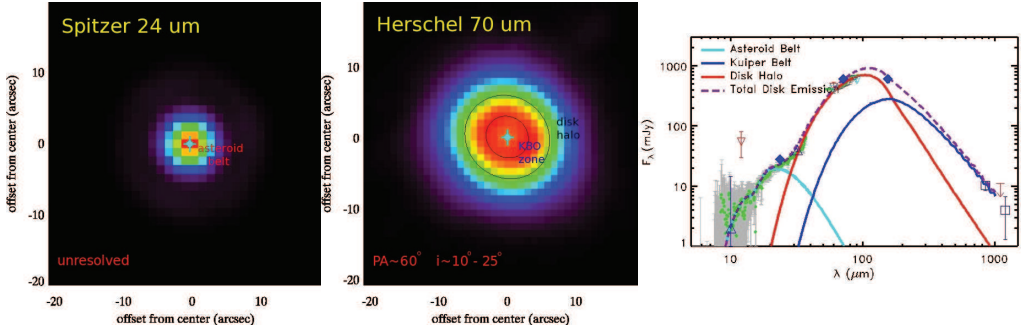} 
 \vspace*{-0.3 cm}
 \caption{Multi-wavelength images of the HR 8799 system along with the disk SED.}
   \label{fig2}
\end{center}
\end{figure}

{\underline{\bf The HR 8799 System}} Among these two-temperature
resolved disks, the most famous one is the HR 8799 system, which has
been under the spotlight since the discovery of the four massive
planets by direct imaging (Marois et al.\ 2010). Besides the massive
planets, the system also has a lot of dust generated from leftover
planetesimals, predominantly at two characteristic dust temperatures:
$\sim$150 K and $\sim$45 K (Su et al.\ 2009). Figure \ref{fig2} shows
the disk images at 24 and 70 $\mu$m along with its SED. At 24 $\mu$m,
the emission is unresolved and dominated by the material closer to
the star. At 70 $\mu$m, the disk is resolved as an elliptical ring
with a position angle of 60\arcdeg, indicating the system is slightly
inclined by 10\arcdeg--25\arcdeg\ from face-on.  From these images and
dust temperatures measured from the SED, we can estimate the radial
distances of these two belts. The four massive planets lie between the
two dust belts as expected if planetary perturbations maintain their
structures and create the dust free zone between them.

{\underline{\bf The Fomalhaut System}} Fomalhaut has a prominent cold
($\sim$50 K) excess, dominated by dust generated in a narrow
planetesimal belt at $\sim$140 AU, well resolved at multiple
wavelengths (e.g.\ Kalas et al.\ 2005; Acke et al.\ 2012; Boley et
al.\ 2012).  The unresolved warm excess near the star was first
discovered in the {\it Spitzer} 24 $\mu$m image by Stapelfeldt et al.\
(2004), and later confirmed by {\it Herschel} at 70 $\mu$m (Acke et
al.\ 2012). Re-analysis of the {\it Spitzer} IRS spectrum centered at
the star confirms that the unresolved excess emission arises from dust
emission at a temperature of $\sim$170 K, an asteroid-belt analog near
the water-ice line (Su et al.\ 2013). Using interferometic
observations, the Fomalhaut system was found to possess a very hot
K-band and hot N-band excesses at the levels of 0.88$\pm$0.12\% and
0.35$\pm$0.10\% above the photosphere (Absil et al.\ 2009; Mennesson
et al.\ 2013). Recently, Lebreton et al.\ (2013) proposed these
interferometric excesses are due to two populations of grains located
at $\sim$0.1--0.3 AU ($\sim$2000 K) and $\sim$2 AU ($\sim$400
K). Although their model can explain the N-band null depths, the model
SED significantly under-estimates the amount of the K-band excess (by
25\%), and over-estimates the excess in the IRS spectrum (i.e.,
$\gtrsim$3 $\sigma$ for wavelengths of 15--30 $\mu$m).

{\underline{\bf The Vega System}} Vega and Fomalhaut are often
referred to as debris disk twins since they are both A-type stars,
located less than 8 pc away, half way through their main-sequence lifetime,
and most importantly, they both have far-infrared excesses first
discovered by {\it IRAS} 30 years ago, arising from dust particles in
an enhanced Kuiper-belt analog. Su et al.\ (2013) present the {\it Spitzer}
IRS spectrum taken at the star position and re-analysis of the {\it
Herschel} 70 $\mu$m image, and suggest the presence of an unresolved
warm excess in the vicinity of the star with a dust temperature of
$\sim$170 K, roughly the temperature at which water ice sublimates. 
With other ancillary data, they confirm that Vega, just
like Fomalhaut, also has an asteroid-belt analog.

{\underline{\bf Other Solar-like Systems}} $\epsilon$ Eri, like
Fomalhaut and Vega, was found to possess a prominent far-infrared
excess revealed by {\it IRAS}. {\it Spitzer} imaging and spectroscopic
data combined with a detailed SED model further suggest a complex
debris structure, with multiple zones in both warm ($\sim$150 K) and
cold ($\sim$50 K) components (Backman et al.\ 2009). Using the same
technique (SED and resolved cold component), the warm and cold
configuration is also seen in the HD 107146 (Hughes et al.\ 2011) and
HD 61005 (Hughes et al., this volume).

\vspace{-0.2cm}
\section{Implications}

In our Solar System, the minor bodies that failed to form planets are
arranged and sculpted by the planets over the course of 4.5 Gyr of
evolution. From knowledge of resonance structures observed in the
Kuiper belt, we know that the outer giant planets have migrated in the
past (e.g., Malhotra 1993) (not formed in-situ). The inner edge of the
Kuiper belt's dusty disk is believed to be maintined by Neptune (Liou
\& Zook 1999), whereas the outer edge of the Asteroid belt is
maintained by the 2:1 orbital resonance with Jupiter (Kirkwood 1867;
Dermott \& Murray 1983). Any debris system with similar structures is
possibly signaling the presence of planets (e.g.\ the HR 8799
system). In the resolved systems discussed above, the orbital ratios
($\sim$10, consistent with the estimated dust temperatures) between
the inner warm and outer cold belts suggest a large gap in the
minor-body population. This large, mostly dust-free zone may be
maintained by one or multiple planet-mass objects through dynamical
interaction, just like the HR 8799 system. For one single planet, the
width of the chaotic zone depends on the location and eccentricity of
the planet and its mass. Therefore, one can use the boundaries of the
warm and cold dust belts to estimate the mass of the possible
perturber.

We use the Vega system as an example where the warm belt is estimated
at $\sim11-14$ AU (Su et al.\ 2013) and the cold belt is at
$\sim$90--120 AU (Su et al.\ 2005, excluding the disk halo). For one
single object on a circular orbit, the mass of the object has to be
greater than 100 $M_J$ (a brown dwarf), which should have
been found in the intensive direct imaging searches (planets with
masses $\gtrsim$3--4 $M_J$ can be ruled out in the 20--70 AU
range, Marois et al.\ 2006). For one single object on an eccentric
orbit, the boundaries of the belts give an expected eccentricity of
0.8 (i.e., warm belt as the pericenter and cold belt as the
apocenter). An object with such a high eccentricity inside the cold
belt will have significant dynamical effects in the system,
forcing the leftover planetesimals to be on highly eccentric orbits and
creating an offset eccentric ring and/or resonance structures, 
which is not what is seen in the resolved disk images (Su et al.\ 2013).

The giant planets are expected to form more efficiently outside the
water-ice line, and these giant planets, once formed, likely
experience inward or outward migration. The discovery of hot Jupiter
systems in the early success of exoplanet searches is the best
evidence for inward migration. Furthermore, if a giant planet did
migrate inward, it is likely to destroy other bodies during the course
of migration, consistent with the low incidence of warm excesses in
such systems (Morales et al.\ 2012). Therefore, the existence of
terrestrial planets and Asteroid belt in our Solar System is probably
because it does not have a hot Jupiter. The same logic applies to
these two-belt systems like Vega, Fomalhaut, $\epsilon$ Eri and HR
8799, having an asteroid belt might indicate a greater chance to
harbor terrestrial planets compared to systems that do not have an
asteroid belt.

\end{document}